\documentclass[a4paper]{article}

\usepackage{subcaption}
\usepackage{xcolor}
\definecolor{lightgray}{gray}{0.75}
\usepackage{INTERSPEECH2021}

\title{Using Self-Supervised Feature Extractors with Attention for Automatic COVID-19 Detection from Speech}
\name{John Mendonça$^{1,2}$, Rubén Solera-Ureña$^{1}$, Alberto Abad$^{1,2}$, Isabel Trancoso $^{1,2}$}

\address{
  $^1$INESC-ID, Portugal\\
  $^2$Instituto Superior Técnico, Universidade de Lisboa, Portugal}
\email{\{john.mendonca, alberto.abad, isabel.trancoso\}@tecnico.ulsiboa.pt, rsolera@hlt.inesc-id.pt}

\begin{document}

\maketitle
\begin{abstract}

The ComParE 2021 COVID-19 Speech Sub-challenge provides a test-bed for the evaluation of automatic detectors of COVID-19 from speech. Such models can be of value by providing test triaging capabilities to health authorities, working alongside traditional testing methods. Herein, we leverage the usage of pre-trained, problem agnostic, speech representations and evaluate their use for this task. We compare the obtained results against a CNN architecture trained from scratch and traditional frequency-domain representations. We also evaluate the usage of Self-Attention Pooling as an utterance-level information aggregation method. Experimental results demonstrate that models trained on features extracted from self-supervised models perform similarly or outperform fully-supervised models and models based on handcrafted features. Our best model improves the Unweighted Average Recall (UAR) from 69.0\% to 72.3\% on a development set comprised of only full-band examples and achieves 64.4\% on the test set. Furthermore, we study where the network is attending, attempting to draw some conclusions regarding its explainability. In this relatively small dataset, we find the network attends especially to vowels and aspirates.

\end{abstract}

\section{Introduction}

In the wake of the global COVID-19 pandemic, extensive research power has been allocated to the development of reliable and cost-effective methods of diagnosis. Current methods rely on intrusive and in-person collections of samples, which may increase the risk of exposure to the virus. Furthermore, given the nature of such tests, their scalability may prove limited. As such, the automatic detection of COVID-19 from speech can assist in mass-testing by providing preliminary test results that do not require expert manpower, and can be performed without the need of leaving one's house, thus maintaining social distancing. Several investigations based on different datasets have appeared during the last months, with varied and inconclusive preliminary results. Initiatives such as the ComParE 2021 COVID-19 Speech Sub-challenge (CSS) \cite{Schuller21-TI2} provide researchers from all over the world a valuable space to share ideas and findings and to compare their results in a common test-bed.

Early attempts at detecting speech-related diseases typically involved cumbersome feature engineering. Unlike mainstream Machine Learning tasks such as Speaker Recognition or ASR (Automatic Speech Recognition), the development of DNN (Deep Neural Network) -based architectures for paralinguistic tasks has an added limitation that data is frequently scarce and unbalanced, which is often an impediment to outperforming results from hand-crafted features \cite{wagner2018deep}. As a solution, researchers typically resort to either data augmentation techniques \cite{yeh2019using, Illium2020}; or by pre-training models on larger datasets, and using them as feature extractors \cite{milde2015using} or by fine-tuning the original model \cite{das2015cross}.

Current research on automatic detection of COVID-19 from speech or respiratory sounds is based on previous work that show how these are differently affected by distinct respiratory diseases and can thus be used to detect cold, asthma, pneumonia, tuberculosis, among others \cite{Gosztolya2017}. These results have inspired several ongoing research projects with a machine learning approach to the problem of automatic detection of COVID-19. The imperative need for properly labeled datasets has been partially addressed by several initiatives, such as those conducted by the University of Cambridge \cite{Chloe2020,Han2021} and the Indian Institute of Science, Bangalore \cite{Sharma2020}, besides our own on-going efforts at INESC-ID/IST-University of Lisbon\footnote{https://www.inesc-id.pt/covid19}.

In \cite{Han2021}, the authors leveraged reported symptoms and perform feature-level and decision-level fusion with the OPENSMILE feature. They reported an AUC (Area Under Curve) of 0.79 in a subset of data crowdsourced from a mobile app. Pinkas et. al \cite{9205643} proposed a three-stage architecture comprising of (1) embeddings extracted from Mockingjay \cite{9054458}; (2) a Recurrent Neural Network that produces specialised sub-models for classification; (3) an ensemble stacking to fuse predictions. They reported a Recall of 78\% on their self-collected dataset.

Attention mechanisms are often used to present a certain degree of explainability to models \cite{serrano2019attention}: attention provides a distribution over attended-to input units, and this is intuitively presented as portraying the relative importance of inputs. However, their ability to provide transparency for model predictions has been questioned \cite{jain-wallace-2019-attention}, namely due to its inconsistency with other feature-importance measures, as well as lack of consistent outputs on multiple runs given the same prediction. Such conclusions have been refuted on the basis that existence does not entail exclusivity \cite{wiegreffe-pinter-2019-attention}. Similar to the work conducted in \cite{MacIntyre2020}, the attention importance weights visualisations of this paper are averaged across multiple trials.

Our research contributions are two-fold: (1) we demonstrate the use of transfer-learning methods based on self-supervised speech representations for COVID-19 detection. More specifically, we show that aggregating these features and combining them with a Fully Connected Network results in comparable performance to handcrafted features and DNN models trained exclusively for this task; (2) we provide a qualitative analysis of the attention importance weights and evaluate this information as a path towards explanation.

This paper is organised as follows: Section 2 describes the architecture employed, namely the Feature Extractors and Feature Aggregators. Section 3 introduces the methodology used for the experiments, and includes a description of the COVID-19 dataset. Section 4 presents the results of our experiments and the qualitative analysis of the predictions. Section 5 draws conclusions and presents directions for future work.

\section{Proposed Methods}

Our proposed architecture leverages self-supervised feature extractors for COVID-19 prediction. As such, our method is divided into 2 steps: 1) Feature Extraction and 2) Final Prediction using a Feed Forward Network with Feature Pooling. We use S3PRL \cite{S3PRL}, a PyTorch based toolkit, for rapid prototyping with different feature extractors.

\subsection{Feature Extraction}

For feature extraction, we extract speech representations every 10ms, with window length set to 25ms. After extraction, the feature is fed through a Fully Connected Layer paired with a $Tanh$, outputting a hidden representation of size $k$. 

\vspace{-0.2cm}
\subsubsection{Spectral Features}

\textbf{Spectrogram} a short-time Fourier transform (STFT) with $n_{fft}=512$ is used to obtain the Raw Spectrogram of the speech signal, resulting in a feature vector of size 257 (including raw energy); \textbf{Mel} the Mel spectrogram is obtained with $n_{freq}= 201$ and 80 mel bins; \textbf{MFCC} the first 13 Mel-Frequency Cepstral Coefficients (MFCCs) are computed together with first and second order derivatives, resulting in a feature vector of size 39; \textbf{FBank} Log Mel-Filter Bank coefficients are extracted using 80 mel bins, together with the first and second order derivatives, resulting in a feature vector of size 240.

\vspace{-0.2cm}
\subsubsection{CPC}

Contrastive Predictive Coding (CPC) \cite{oord2018representation} is an approach for unsupervised learning from high-dimensional data by translating a generative modelling problem to a classification problem. That is, it combines predicting future observations (predictive coding) with a probabilistic contrastive loss. This allows for the extraction of representations that are useful for phone and speaker recognition tasks. Although we did not find any previous work on the use of CPC as a problem-agnostic feature extractor, contrastive loss as an objective for speech embeddings has been used for emotion recognition \cite{nandan2020language}. We extract from the hidden state of the encoder network feature vectors of dimension 256.

\vspace{-0.2cm}
\subsubsection{PASE}

The Problem-Agnostic Speech Encoder (PASE) \cite{9053569} encodes the raw speech waveform into a representation that is fed to multiple regressors and discriminators (called workers) to produce meaningful and robust representations. The worker architecture receives encoded representation to solve seven self-supervised tasks (including reconstruction of waveform in an autoencoder fashion, predicting the Log power spectrum, MFCC and prosodic features). The authors evaluate PASE on speaker identification, emotion classification, and ASR. For this task we use PASE+, an improved version of PASE for noisy and reverberant environments, to extract vectors of dimension 256 from the output of the encoder.

\vspace{-0.2cm}
\subsubsection{TERA}

TERA, which stands for Transformer Encoder Representations from Alteration \cite{Liu2020TERASL}, uses a multi-target auxiliary task to pre-train Transformer Encoders on a large amount of unlabelled speech. The authors evaluate TERA on phone classification, speaker recognition, and speech recognition, but its self-supervised nature makes extraction of speech representations or fine-tuning with downstream models possible. To the best of our knowledge, no previous research on TERA for paralinguistic tasks exists. We extract the hidden states of the last Transformer Encoder layer, with dimension 768 as features.

\vspace{-0.2cm}
\subsubsection{Mockingjay}

Mockingjay is a Bidirectional Transformer architecture that is designed to predict the current frame by conditioning on both past and future contexts \cite{9054458}. Mockingjay can be considered equivalent to TERA, when TERA is only using the time objective. The authors evaluate Mockingjay on sentiment analysis tasks by extracting a feature representation that consists of the weighted sum of the Transformer hidden representations (denoted \textit{LARGE-WS}). In our experiments, we also use \textit{LARGE-WS}, which has a dimension of 768.

\subsection{Feature Pooling and Prediction}

The feature extraction step outputs speech vector representations at the frame level (10ms stride). For prediction, an utterance level representation is required. We hypothesise that COVID-19, such as other respiratory diseases, affects distinct parts of the speech production system differently, and may present itself more distinctively in certain speech sounds. To this end, we evaluate the inclusion of a Self-Attention Pooling Layer \cite{Safari2020}. This layer is a dot product attention where the keys and the values correspond to the same representation and the query is only a trainable parameter. The utterance-level representation can be obtained as follows:
\begin{equation}
    H = softmax(W_cX^T)X
\end{equation}
where $X=[x_1,x_2,...,x_T] \in \mathbb{R}^{T\times k} $ is the sequence of features of size $k$ and $W_C \in \mathbb{R}^{k}$ is a trainable parameter.

The choice of attention pooling as a frame weighting mechanism has been successfully applied to whisper detection \cite{naini2020whisper} and speaker recognition \cite{Safari2020}, which followed similar motivations to ours. We compare results obtained with Self-Attention Pooling with a simple Mean pooling layer.

Before prediction, the pooled utterance-level representation goes through a Dropout layer and Fully Connected Layer to produce a two-dimensional output representing unnormalized logits. We experimented different approaches including adding a Regularisation layer (with and without Dropout) but found performance degraded with these changes. 

\subsection{CNN + SAP}

As a comparative method, we train a DNN in a fully-supervised manner. The architecture is a three layer CNN (Convolutional Neural Network) which tries to find a robust representation of the original signal. It is comprised of a series of convolutions followed by an activation (ReLU) and a Dropout layer The architecture receives as input the spectrogram from raw audio signal (as presented in 2.1.1). Before being fed to the convolutional network, the spectrogram is averaged pooled with a window and stride of 50ms. The size of the kernels and the number of channels produced by the convolutional layers are 5 and 160, respectively. Stride and padding are set to 1 and 2, respectively. The last convolutional layer outputs speech representations of hidden-size 160 that are aggregated using Self-Attention Pooling. Before prediction, the pooled representation is passed through a Feed-Forward Network, consisting of a Linear layer with an output of size 160, followed by a ReLU. A final Linear Layer produces a two-dimensional vector of unnormalized logits. Considering this model is trained from scratch, we augment the training dataset using WavAugment \cite{kharitonov:hal-03070321}. The augmentation, which receives as input the raw wave representation, consists of pitch randomization, reverberance and time clipping, in that order. This augmentation doubles the original training set.

\section{Experimental Set-up}

We ran our experiments for 10,000 steps, with evaluation every 200 steps. For all models we train using \textit{CrossEntropyLoss} as the criterion. Optimisation is done with \textit{AdamW} \cite{loshchilov2018decoupled}, with a learning rate of 0.0004. For the CNN training, we schedule a learning rate that decreases linearly from an initial learning rate of 0.0002 to 0, after a warm-up period of 1400 steps. Batch size was set to 8 due to hardware limitations. We present results with the size of the hidden representation $k$ set to \{128, 256, 512, 768\}.

\subsection{COVID-19 SPEECH (C19S) corpus}

The COVID-19 SPEECH (C19S) corpus is a curated subset of the Cambridge COVID-19 Sound database \cite{Chloe2020,Han2021}, a worldwide-crowdsourced corpus with examples of breathing, coughs and speech recorded "in-the-wild". The C19S corpus contains 893 speech recordings from 366 participants and their corresponding self-reported COVID-19 status labels (positive/negative), distributed in three speaker-independent and gender-balanced subsets: Train (72 COVID-19 positives and 243 negatives), Development (142 positives and 153 negatives) and a blind Test set (283 samples). All the recordings are in PCM format, single-channel, 16 bits per sample, sampling rate of 16 kHz, and were normalised in amplitude. In each recording, participants recorded themselves reading a given prompt (1-3 times).

In our preliminary analysis of the C19S corpus, we noticed some files present a reduced bandwidth of 4 kHz, hypothetically corresponding to audio samples originally recorded at a sampling rate of 8 kHz. Namely, 16, 13 and 7 narrow-band files were detected in the Train, Development and Test subsets, respectively, all of them corresponding to the COVID-19 positive class \footnote{Similar issues were found in the cough (C19C) corpus.}. From our analysis of the provided baselines and our own systems, we consider that this issue could be affecting their performance by making classifiers pay attention to this spurious condition during training. For this reason, we decided to remove all the narrow-band recordings in the original Train and Development subsets, even at the cost of reducing the number of examples in the minority positive class. The resulting dataset contains 299 samples in the Train subset (56 positives) and 282 samples in the Development subset (129 positives). The Test subset is kept untouched so as to stick to the original definition and evaluation conditions of the ComParE 2021 CSS Sub-challenge. For the rest of the paper, all decisions regarding our proposal are taken based on development results on the C19C-reduced dataset, denoted $dev_{fband}$, although results with the original C19S dataset ($dev$) are included for
comparison.

\section{Results}

\begin{figure*}[t]
\centering
\begin{subfigure}{.49\textwidth}
  \centering
  \includegraphics[width=1.02\linewidth]{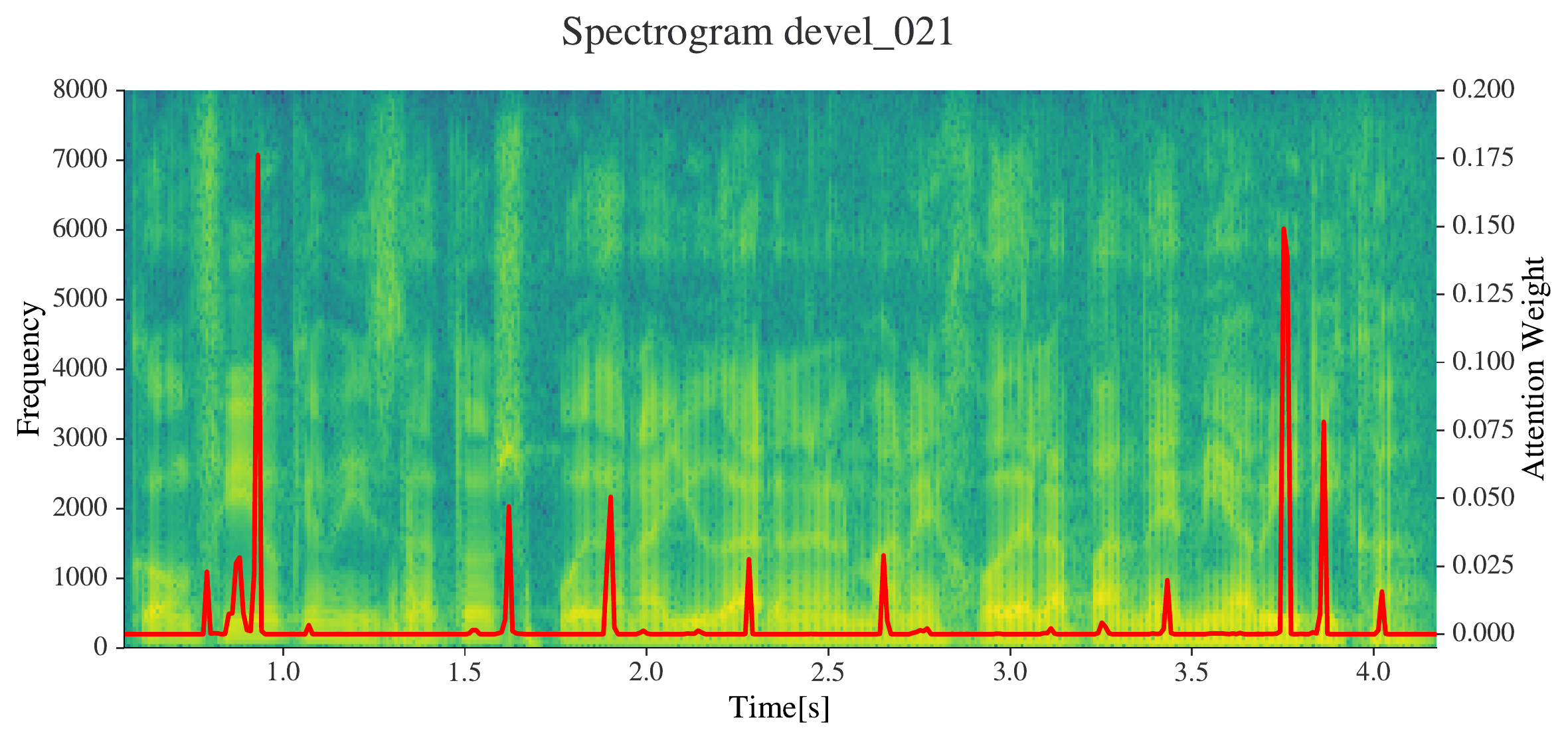}
  \caption{Spectrogram of sentence: "Confío en que mis datos puedan ayudar en el manejo de la pandemia vírica.", repeated once.}
  \label{fig:att1}
\end{subfigure}\hfill
\begin{subfigure}{.49\textwidth}
  \centering
  \includegraphics[width=1.02\linewidth]{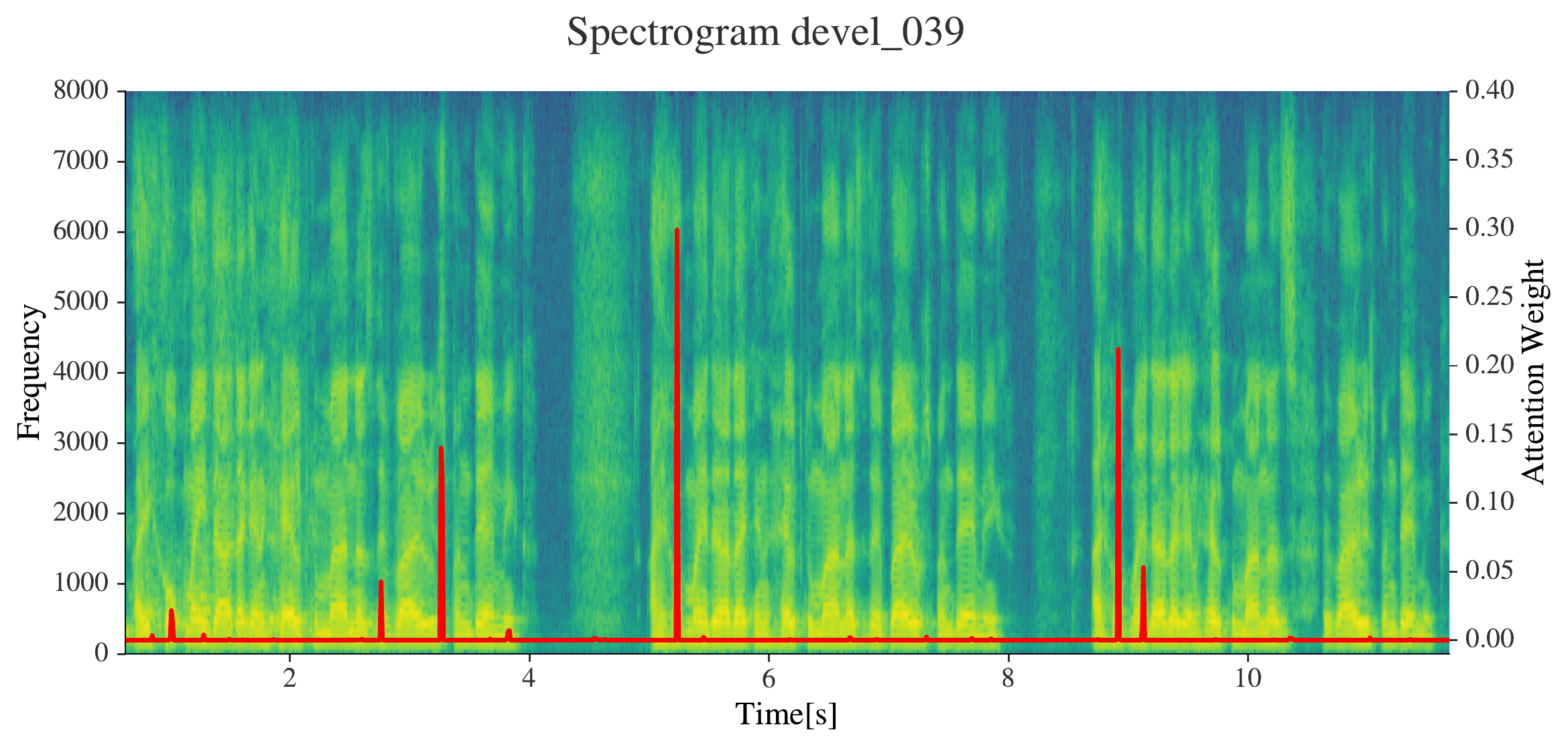}
  \caption{Spectrogram of sentence: "I hope my data can help manage the virus pandemic.", repeated 3 times by a non-native speaker.}
  \label{fig:att2}
\end{subfigure}
\vspace{-0.2cm}
\caption{Attention importance weights (red) plotted against the spectrogram of correctly classified Covid-19 positive subjects. The attention weights are an average across 5 trials and correspond to the best performing self-supervised model using SAP of $dev_{fband}$.}
\label{fig:att}
\vspace{-0.3cm}
\end{figure*}

The results for the models with the best development performance during training in terms of UAR (Unweighted Average Recall) 
are summarised in Table \ref{tab:res}. We include the baseline methods \cite{opensmile, xbow, amiriparian2017Snore, 10.5555/3122009.3242030, tzirakis2018end2you} for comparison. The best performing models for each class are marked in bold, while the best overall model for each subset is highlighted in greyscale. We submit to $test$ our best performers in $dev_{fband}$. The presented results were obtained using the best $k$ for that feature. Our proposed self-supervised features show competitive results when compared to the baseline approaches in both $dev$ and $dev_{fband}$. In fact, FBank-Mean, CPC-SAP and PASE+-SAP outperform the best baseline model by at least 1.8\% UAR in $dev$. Meanwhile, TERA and Mockingjay are the worse performers, with similar performances to the ones of Spectral Features. This indicates the Transformer Encoder architecture using time, channel and magnitude reconstruction pre-training objectives fails to produce meaningful representations for COVID-19 detection, when compared to CPC and PASE+.

\begin{table}[h]
\caption{Performance results (Unweighted Average Recall-UAR) on the COVID-19 SPEECH (C19S) corpus.}
\label{tab:res}
\centering
\begin{tabular}{cccccc}
\multicolumn{1}{c|}{\textbf{Feature}} & \multicolumn{2}{c|}{$dev$} & \multicolumn{2}{c|}{$dev_{fband}$} & $test$ \\ \hline
\multicolumn{6}{c}{\textbf{Baseline Approaches}}                                                                                                                \\ \hline
\multicolumn{1}{c|}{openSMILE}        & \multicolumn{2}{c|}{57.9}                  & \multicolumn{2}{c|}{58.4}                        & \fcolorbox{lightgray}{lightgray}{\textbf{72.1}}                   \\
\multicolumn{1}{c|}{openXBOW}         & \multicolumn{2}{c|}{66.3}                  & \multicolumn{2}{c|}{59.3}                            & 68.7                   \\
\multicolumn{1}{c|}{deepSPEC}     & \multicolumn{2}{c|}{56.0}                  & \multicolumn{2}{c|}{51.6}                        & 60.4                   \\
\multicolumn{1}{c|}{auDEEP}           & \multicolumn{2}{c|}{62.2}                  & \multicolumn{2}{c|}{58.8}                        & 64.2                   \\
\multicolumn{1}{c|}{End2You }          & \multicolumn{2}{c|}{\textbf{70.5}}                  & \multicolumn{2}{c|}{\textbf{69.0}}                            & 68.7                   \\ \hline
\multicolumn{4}{c}{}                                 \\ \hline
\multicolumn{1}{c|}{Aggregation}      & Mean      & \multicolumn{1}{c|}{SAP}       & Mean         & \multicolumn{1}{c|}{SAP}           &                        \\ \hline
\multicolumn{6}{c}{\textbf{Spectral Features}}                                                                                                                  \\ \hline
\multicolumn{1}{c|}{Spectogram}       & 58.7      & \multicolumn{1}{c|}{63.2}      & 55.5         & \multicolumn{1}{c|}{57.6}          &  -                      \\
\multicolumn{1}{c|}{Mel}              & 61.4      & \multicolumn{1}{c|}{64.7}      & 56.5         & \multicolumn{1}{c|}{59.5}          &  -                      \\
\multicolumn{1}{c|}{MFCC}             & 64.7      & \multicolumn{1}{c|}{63.6}      & 54.3         & \multicolumn{1}{c|}{62.7}          &  -                      \\
\multicolumn{1}{c|}{FBank}            &\fcolorbox{lightgray}{lightgray}{\textbf{72.5}}      & \multicolumn{1}{c|}{63.9}      & \textbf{64.3}         & \multicolumn{1}{c|}{63.0}         &   -                     \\ \hline
\multicolumn{6}{c}{\textbf{Self-Supervised Features}}                                                                                                           \\ \hline
\multicolumn{1}{c|}{CPC}              & 62.3      & \multicolumn{1}{c|}{\textbf{72.1}}      & 63.8         & \multicolumn{1}{c|}{68.2}          & 56.7                       \\
\multicolumn{1}{c|}{PASE+}             & 70.7      & \multicolumn{1}{c|}{71.8}      & \textbf{69.8}         & \multicolumn{1}{c|}{66.4}          & \textbf{58.6}                       \\
\multicolumn{1}{c|}{TERA}             & 59.7      & \multicolumn{1}{c|}{61.3}      & 58.8         & \multicolumn{1}{c|}{57.3}          &  -                      \\
\multicolumn{1}{c|}{Mockingjay}         & 60.0      & \multicolumn{1}{c|}{58.2}      & 55.9         & \multicolumn{1}{c|}{56.7}          &  -                      \\ \hline
\multicolumn{6}{c}{\textbf{Supervised Approaches}}                                                                                                                \\ \hline
\multicolumn{1}{c|}{CNN}              & -         & \multicolumn{1}{c|}{68.9}         & -            & \multicolumn{1}{c|}{\fcolorbox{lightgray}{lightgray}{72.3}}             &  64.4                      \\
\end{tabular}
\end{table}

The use of Self-Attention Pooling (SAP) fails to bring consistent improvements of results. Our self-supervised features in $dev$ report small improvements of performance when moving from Mean to SAP, but when SAP is used in conjunction with Spectral Features results are mostly worse in $dev$ but better in $dev_{fband}$. CPC appears to benefit the most from SAP, consistently improving performance in both sets (9.8\% and 4.4 \% in $dev$ and $dev_{fband}$, respectively).

We note an overall loss in performance when removing the narrow-band files in $dev_{fband}$, as expected. The spectral features, in particular, report significant drops. Considering these features contain information pertaining bandwidth, we believe these models were using band-detection as means to classify positive classes. This explains why FBank managed to outperform all other models in $dev$. Surprisingly, our CNN architecture's performance improves with the removal of these files. 

With respect to our model's performance in $test$, we note that there is a drop in performance when compared to results in $dev$, contrary to what happens to the baseline approaches, which report improvements. This can be partially explained by our models being trained on full-band examples only, which reduced the number of COVID-19 positive examples. 

As detailed in Section 2, one of the hyper-parameters of our model is the size of the hidden representation $k$. We present the performance changes on $dev_{fband}$ in Table \ref{tab:N_table}. Overall performance metrics remain consistent through the different values of $k$. Although minor, TERA and Mockingjay report improvements when bottlenecking, which can be due to the fact these features are obtained by concatenating several hidden representations, and thus might contain information overlap.

\begin{table}[h]
\caption{Performance Results (UAR) using Self-Supervised Features on $dev_{fband}$ with varying $k$.}
\vspace{-0.2cm}
\centering
\label{tab:N_table}
\begin{tabular}{ccccl}
\multicolumn{1}{l}{}         & \multicolumn{4}{c}{\textbf{$k$}}     \\ \cline{2-5} 
\multicolumn{1}{c|}{\textbf{Pooling}} & 128  & 256  & 512  & 768  \\ \hline
\multicolumn{5}{c}{\textbf{CPC}}                                  \\ \hline
\multicolumn{1}{c|}{Mean}    & 62.2 & \textbf{63.8} & 60.3 & 61.2 \\
\multicolumn{1}{c|}{SAP}     & 60.8 & \textbf{68.2} & 64.0 & 61.7 \\ \hline
\multicolumn{5}{c}{\textbf{PASE+}}                                \\ \hline
\multicolumn{1}{c|}{Mean}    & 66.2 & 65.3 & \textbf{69.9} & 65.5 \\
\multicolumn{1}{c|}{SAP}     & \textbf{66.4} & 64.8 & 65.2 & 59.3 \\ \hline
\multicolumn{5}{c}{\textbf{TERA}}                                 \\ \hline
\multicolumn{1}{c|}{Mean}    & 57.5 & 57.5 & \textbf{58.8} & 58.4 \\
\multicolumn{1}{c|}{SAP}     & 56.4 & 55.8 & \textbf{57.3} & 57.4 \\ \hline
\multicolumn{5}{c}{\textbf{Mockingjay}}                           \\ \hline
\multicolumn{1}{c|}{Mean}    & \textbf{55.9} & 54.1 & 55.2 & 54.1 \\
\multicolumn{1}{c|}{SAP}     & \textbf{56.7} & 55.6 & 55.6 & 54.5
\end{tabular}
\end{table}

\vspace{-0.5cm}
\subsection{Attention: A Qualitative Evaluation}

As a qualitative examination of the network's classification capabilities, we present average attention importance weights across 5 trials in Figure \ref{fig:att}. It can be observed that the network focuses itself on certain points of the spectrogram in both examples. 

In Figure \ref{fig:att1}, the first segment with high attention activity ([0-1]s) corresponds to the Spanish word \textit{"confío"}. Maximum attention is paid to the transition between the last two phones. The second segment ([3.5-4]s) also shows maximum attention to the transition between the last two vowels in \textit{"pandemia"} followed by the first vowel of \textit{"vírica"}, thus seeming to privilege the vowel /i/.
The remaining smaller peaks also correspond to 
other vowel transitions and nasal sounds. Regarding Figure \ref{fig:att2}, in [2-4]s, the attention weights present themselves maximally around the last vowel of \textit{"virus"} and the first nasal sound of \textit{"pandemic"}; in [4-6]s and [8.10]s the peaks correspond to the transition between the aspirate /h/ and the vowel (repetition of \textit{"hope"}).

The attention importance weights in both examples point towards the network privileging vowels and aspirates, in English, for detecting COVID-19. The idea that COVID presents itself in vowel sounds is aligned with previous research \cite{Quatieri2020}. More specifically, COVID-19 disrupts the entrainment of the vocal folds during phonation, especially for /i/ \cite{ismail2020detection}. Aspirates deserve a more detailed analysis with more examples, but we hypothesise these sounds also provide informative cues due to its similarities with coughs and expirations. 

\section{Conclusions}

In this work, we proposed using Self-Supervised feature extractors for the task of COVID-19 prediction from speech. Results show that PASE+ and CPC features are able to perform comparably, or even better, to baseline approaches in the development set. Moreover, we employed attention mechanisms as a feature aggregation method and performed a qualitative analysis of the attention importance weights to draw some intuition behind the predictions. Our analysis indicate the network privileges features extracted from vowels and aspirates for prediction. However, it is not clear whether models based on these sounds alone are able to, individually, outperform ones based on full sentences. 

One of the main concerns pertaining automatic detection of COVID-19 from speech is the performance inconsistencies across datasets. 
As future work, we plan to address this problem using different datasets, including our own COVID-19 crowdsourced dataset, which is still under construction. Additionally, we will assess predictions based only on sustained vowels, coughs and read/spontaneous speech to better understand individual contributions, and work towards a more succinct COVID test. An interesting research direction is to bridge this work with previous research on breathing event detection \cite{Mendonca2020} for this task.

\section{Acknowledgements}

This work has been supported by national funds through Fundação para a Ciência e a Tecnologia (FCT), under project UIDB/50021/2020, and by FEDER, Programa Operacional Regional de Lisboa, Agência Nacional de Inovação and CMU Portugal, under grant BI$\mid$2020/091.

\bibliographystyle{IEEEtran}

\bibliography{mybib}

\end{document}